\documentclass[12pt]{article}

\pdfoutput = 1

\usepackage{amsmath}
\usepackage{graphicx}
\usepackage{subcaption}
\usepackage{titlesec}
\usepackage{bigints}
\usepackage{xcolor}
\usepackage{hyperref}

\hypersetup{colorlinks, linkcolor={blue}, citecolor={red}}
\titlelabel{\thetitle.\quad}
\numberwithin{equation}{section}

\begin{document}

\date{today}
\title{{\bf{\Large Meissner like effect in holographic superconductors with back reaction}}}
\author{
{\bf {\normalsize Suchetana Pal}$^{a}
$\thanks{suchetanapal92@gmail.com, sp15rs004@iiserkol.ac.in}}
{\bf {\normalsize , Saumya Ghosh}$^{b}
$\thanks{sgsgsaumya@gmail.com, sg14ip041@iiserkol.ac.in}}
{\bf {\normalsize , Sunandan Gangopadhyay}$^{c}
$\thanks{sunandan.gangopadhyay@gmail.com, sunandan.gangopadhyay@bose.res.in}}\\
$^{a}${\normalsize\textit{Department of Physical Sciences,}}\\
{\normalsize\textit{Indian Institute of Science Education and Research Kolkata,}}\\
{\normalsize\textit{Mohanpur, Nadia, West Bengal, 741 246, India.}}\\
$^{c}${\normalsize\textit{Department of Theoretical Sciences,}}\\
{\normalsize\textit{S. N. Bose National Centre for Basic Sciences,}}\\
{\normalsize\textit{Block - JD, Sector - III, Salt Lake, Kolkata - 700 106, India.}}}
\date{}

\maketitle

\begin{abstract}
\noindent In this article we employ the matching method to analytically investigate the properties of holographic superconductors in the framework of Maxwell electrodynamics taking into account the effects of back reaction on spacetime. The relationship between the critical temperature ($T_{c}$) and the charge density ($\rho$) has been obtained first. The influence of back reaction on Meissner like effect in this holographic superconductor is then studied. The results for the critical temperature indicate that the condensation gets harder to form when we include the effect of back reaction. The expression for the critical magnetic field ($B_{c}$) above which the superconducting phase vanishes is next obtained. It is observed from our investigation that the ratio of $B_{c}$ and $T_{c}^{2}$ increases with the increase in the back reaction parameter. However, the critical magnetic field $B_c$ decreases with increase in the back reaction parameter.
\end{abstract}

\section{Introduction}
The AdS/CFT correspondence which gives a sound realization of the holographic principle, fundamentally originated from superstring theory \cite{Maldacena, Witten}. The duality claims the equivalence of a strongly coupled $D$-dimensional gauge theory with a gravitational theory in $(D+1)$-dimensional anti-deSitter (AdS) spacetime. In other words this duality states the equivalence between two theories one of which is strongly coupled and the other is weakly coupled. For the past two decades many successful connections between condensed matter physics and gravitational theories have been built using this AdS/CFT duality, also known as the gauge/gravity correspondence in the literature. To describe finite temperature field theories, the gravitational part is replaced by AdS black holes. More specifically, exploiting the gauge/gravity duality one can use general relativity as a tool to describe various strongly correlated systems of condensed matter physics. One of the phenomena in condensed matter system that is explained by this correspondence is the phenomena of superconductivity \cite{Hartnoll 01, Hartnoll et. al. 01}.  

Holographic superconductors have been studied extensively in the last few years and the results obtained on the field theory side from the holographic superconducting gravitational models showed considerable promise to explain some of the glaring features of high $T_c$ superconductors. The basic idea involved in holographic superconductors is the following. The local spontaneous $U(1)$ symmetry breaking of a charged black hole, minimally coupled to a complex scalar field, was first studied in \cite{Gubser 01, Gubser 02, Gubser 03}. The gravitational theory in the bulk can be mapped into a dual field theory, residing on the boundary of the AdS spacetime, using the AdS/CFT correspondence. This boundary theory suffers a global $U(1)$ symmetry breaking. In the last few years investigation on holographic superconductors have been widely explored in the context of Maxwell electrodynamics \cite{Hartnoll et. al. 02}-\cite{Gangopadhyay et. al. 03} as well as non-linear electrodynamics \cite{Jing et. al. 01}-\cite{Roychowdhury 01}.

It is also well known that superconductors expel magnetic fields as the temperature is lowered below a critical temperature ($T_c$). In the presence
of an external magnetic field, ordinary superconductors can be classified into two classes, namely type I and type II. Using the AdS/CFT dictionary it has been found that at low temperatures ($T < T_c$), $s$-wave condensate in holographic superconductors expels the magnetic field \cite{Maeda et. al. 01, Montull et. al. 01}. Such studies were first carried out in \cite{Albash et. al. 01, Albash et. al. 02}. Thereafter studies of Meissner like effect in holographic superconductors have also been carried out using the matching method \cite{Gregory et. al. 01} in \cite{Gangopadhyay 01, Suchetana-Annals}.

In this paper we have investigated the influence of back reaction on holographic superconductors as well as the Meissner like effect in holographic superconductors. Our holographic superconducting theory consists of a Einstein-Hilbert gravity theory along with a complex scalar field minimally coupled to Maxwell field. The model also involves the effect of back reaction of matter fields on the bulk spacetime. Hence, we are away from the probe limit through out the entire analysis. Studies involving back reaction have been carried out earlier in \cite{Gangopadhyay 02, Debabrata-EPJC}.

The paper is organized as follows. In section 2, the basic formalism for the $d$-dimensional holographic superconductor considering the effect of back reaction in the spacetime geometry is presented. In section 3, we obtain the relationship between the critical temperature and the charge density using the matching method approach. In section 4, we investigate the Meissner like effect using the same approach. 
We finally conclude in section 5.

\section{Basic formalism}
The action of the gravitational dual able to describe the phase transition in the conformal field theory living in the boundary reads 

\begin{eqnarray}
\label{action}
\mathcal{S}&=& \int d^dx~\frac{\sqrt{|g|}}{2\kappa^2}~\Big[R-2\Lambda+2\kappa^2\mathcal{L}_{m}\Big]~,~~~~~~\Lambda=-\frac{(d-1)(d-2)}{2L^2}~,\\
\mathcal{L}_m&=&-\frac{1}{4}F^{\mu\nu}F_{\mu\nu}-\Bigg[(D_\mu\psi)^*D^\mu\psi+m^2\psi^*\psi\Bigg] \nonumber\\
&\equiv&\mathcal{L}_1-\mathcal{L}_2~;~~~\mu,\nu=0,1,2,...d~.\label{Lm}
\end{eqnarray}  
It is basically the Einstein-Hilbert gravity theory with a complex scalar field minimally coupled to the Maxwell field. Where the symbols have their usual meanings and $\kappa^2=8\pi G_{d}$, with $G_d$ being the $d$ dimensional Newton's gravitational constant. $\mathcal{L}_m$ denotes the Lagrangian density of the matter sector. The second term ($\mathcal{L}_2$) in the above equation (\ref{Lm})  consists of the complex scalar field $\psi$, the covariant derivative $D_\mu$ is defined as $D_\mu\equiv \partial_\mu-iqA_\mu$ where $A_\mu$ is the gauge field. On the other hand the first term ($\mathcal{L}_1$) gives the dynamics of the gauge field. 
Next we obtain the equations of motion by varying the action (\ref{action}) with respect to the field variables. The equations of motion for the field variables $g_{\mu\nu},~A_\mu$ and $\psi$, respectively, are given as follows

\begin{subequations}
\label{EoM}
\begin{eqnarray}
\label{Metric EoM}
\delta_{g_{\mu\nu}}&:&\Bigg[R_{\mu\nu}-\frac{1}{2}Rg_{\mu\nu}+\Lambda g_{\mu\nu}\Bigg]-2\kappa^2\Bigg[\frac{1}{2}\mathcal{L}_1\,g_{\mu\nu}+\frac{1}{2}F_{\mu\beta}{F_{\nu}}^\beta-\frac{1}{2}\mathcal{L}_2\,g_{\mu\nu} ~~~~~~~~~~~~~~\nonumber\\
&\qquad&~~~~~+(D_\mu\psi)^*(D_\nu\psi)\Bigg]=0  \\
\label{Gauge EoM}
\delta_{A_\mu}&:&\partial_\nu\Bigg[\sqrt{|g|}F^{\nu\mu}\Bigg]-2\sqrt{|g|}q^2A^\mu\psi^*\psi=0         \\
\label{Scalar EoM}
\delta_{\psi}&:&\partial_\mu\left[\sqrt{|g|}\partial^\mu\psi\right]-iq\sqrt{|g|}A^\mu\partial_\mu\psi-iq\partial_\mu\left[\sqrt{|g|}A^\mu\psi\right]-\sqrt{|g|}q^2A^\nu A_\nu\psi -\sqrt{|g|}m^2\psi=0~.\nonumber\\
\end{eqnarray}
\end{subequations}
\normalsize
\noindent The $d$-dimensional plane-symmetric black hole metric in $AdS$ spacetime with back reaction reads
\begin{equation}
\label{metric}
ds^2=-f(r)e^{-\chi(r)}dt^2+\frac{1}{f(r)}dr^2+r^2h_{ij}dx^idx^j
\end{equation}
where $h_{ij}dx^idx^j$ is the metric on a $(d-2)$-dimensional hypersurface which has a flat geometry. 

Now we consider a change of coordinate from $r\rightarrow z=\frac{r_+}{r}$ where $r_+$ is the horizon radius ($f(r_+)=0$) and make the following ansatz for the gauge field and scalar field 
\begin{equation}
\label{Gauge-Scalar-ansatz}
A_\mu=(\phi(r),~0,~0,~...),~\psi=\psi(r)~.
\end{equation} 
By putting $\mu=\nu=0$ in eq.(\ref{Metric EoM}), we obtain
\begin{equation}
\label{11}
f'(z)-\frac{d-3}{z}f(z)+\frac{(d-1)r_{+}^{2}}{L^{2}z^{3}}-\frac{2k^{2}}{d-2}\bigg(z\psi^{'2}(z)f(z)+\frac{r_{+}^{2}}{z^{3}}m^{2}\psi(z)+\frac{r_{+}^{2}\phi^{2}(z)\psi^{2}(z)e^{-\chi(z)}}{z^{2}f(z)}+\frac{1}{2}z\phi^{'2}(z)\bigg)=0~.
\end{equation}
Now setting $\mu=\nu=1$ in eq.(\ref{Metric EoM}) and then adding up with eq.(\ref{11}) we obtain
\begin{eqnarray}
\label{00+11}
\chi'(z)-\frac{4\kappa^2 r_+^2}{(d-2)z^3}\Bigg[\frac{z^4}{r_+^2}{\psi'}^2(z)+\frac{\phi^2(z)\psi^2(z)e^{\chi(z)}}{f^2(z)}\Bigg]=0~.
\end{eqnarray}
Also putting $\mu=0$ in eq.(\ref{Gauge EoM}), we obtain
\begin{equation}
\label{1}
\phi''(z)+\Bigg[\frac{\chi'(z)}{2}-\frac{d-4}{z}\Bigg]\phi'(z)-\frac{2r_+^2\phi(z)\psi^2(z)}{f(z)z^4}=0~.
\end{equation}
Finally rewriting eq.(\ref{Scalar EoM}) in terms of $z$ gives
\begin{equation}
\label{Scalar EoM A}
\psi''(z)+\Bigg[\frac{f'(z)}{f(z)}-\frac{d-4}{z}-\frac{\chi'(z)}{2}\Bigg]\psi'(z)+\frac{r_+^2}{z^4}\Bigg[\frac{\phi^2(z)e^{\chi(z)}}{f^2(z)}-\frac{m^2}{f(z)}\Bigg]\psi(z)=0~.
\end{equation}

\section{Critical temperature ($T_c$) in terms of charge density ($\rho$)}
The Hawking temperature of a black hole is given by
\begin{equation}
\label{HTemp}
T_H=\frac{f'(r_+)e^{-\chi(r_+)/2}}{4\pi}~.
\end{equation}
To solve the non-linear equations $(\ref{11})-(\ref{Scalar EoM A})$, we need to set the boundary conditions at the black hole horizon $r=r_+$ and at spatial infinity $r=\infty$. In this context we recall that $f(r=r_+)=0$ and $e^{\-\chi(r=r_+)}$ is finite. We also set $\lim\limits_{r\rightarrow\infty}e^{-\chi(r)}\rightarrow1$. The matter fields obey \cite{Gubser 01, Gubser 02}
\begin{eqnarray}
\label{Bc 1}
\phi(r)&=&\mu-\frac{\rho}{r^{d-3}} \\
\label{Bc2}
\psi(r)&=&\frac{\psi_-}{r^{\Delta_-}}+\frac{\psi_+}{r^{\Delta_+}}
\end{eqnarray}
where
\begin{equation}
\label{Delta+-}
\Delta\pm=\frac{(d-1)\pm\sqrt{(d-1)^2+4m^2L^2}}{2}~.
\end{equation}
Here $\mu$ and $\rho$ are the duals to the chemical potential and charge density of the conformal field theory part. We also choose $\psi_-=0$ so that $\psi_+$ is the expectation value of the condensation operator $J$ at the boundary. For the matter field to be regular we require $\phi(r_+)=0$ and $\psi(r_+)$ to be finite. In terms of the coordinate variable $z$ this reads $\phi(z=1)=0$ and $\psi(z=1)$ to be finite.  

\noindent At the critical temperature $T=T_c$, $\psi(z)=0$. Using this fact in eq.(\ref{00+11}) we have
\begin{equation}
\label{crtcl temp 1}
\chi'(z)=0~.
\end{equation}
With this relation and our previous argument we get 
\begin{equation}
\label{crtcl temp 2}
\lim\limits_{z\rightarrow0}e^{-\chi(z)}\rightarrow1 \implies \chi(z)=0~.
\end{equation} 
Using eq.(s)(\ref{crtcl temp 1}, \ref{crtcl temp 2}) in eq.(\ref{1}), we get the following behavior for the field $\phi$ at the critical temperature
\begin{equation}
\label{crtcl temp phi}
\phi''(z)-\frac{d-4}{z}\phi'(z)=0~. 
\end{equation} 
Now using the boundary condition (\ref{Bc 1}), we solve the above equation to get
\begin{equation}
\label{sol. 1}
\phi(z)=\lambda r_+(1-z^{d-3})
\end{equation}
where
\begin{equation}
\lambda=\frac{\rho}{r_+^{d-2}}~.
\end{equation}
Now the field equation for $f(z)$ (\ref{11}) at the critical temperature $T=T_c$ becomes
\begin{equation}
\label{crtcl temp fr}
f'(z)-\frac{d-3}{z}f(z)+\frac{(d-1)r_+^2}{L^2z^3}-\frac{\kappa^2z}{d-2}{\phi'}^2(z)=0~.
\end{equation}
Substituting the solution of $\phi(z)$ from eq.(\ref{sol. 1}) in the above equation yields
\begin{equation}
\label{crtcl temp fr 1}
f'(z)-\frac{d-3}{z}f(z)+\frac{(d-1)r_+^2}{L^2z^3}-\frac{(d-3)^2\kappa^2\lambda^2r_+^2}{d-2}z^{2d-7}=0~.
\end{equation}
The solution of this equation (\ref{crtcl temp fr 1}) subject to the condition $f(z=1)=0$ reads
\begin{equation}
\label{sol. 3}
f(z)=r_+^2\Bigg[\frac{1}{L^2z^2}-\Big(\frac{1}{L^2}+\frac{d-3}{d-2}\kappa^2\lambda^2\Big)z^{d-3}+\frac{d-3}{d-2}\kappa^2\lambda^2z^{2(d-3)} \Bigg]~.
\end{equation}
In the rest of the analysis we shall work with $L=1$. With this value of $L$ the form of $f(z)$ reduces to
\begin{equation}
\label{sol. 3a}
f(z)=\frac{r_+^2}{z^2}g_0(z)
\end{equation}
where 
\begin{equation}
\label{gz}
g_0(z)=1-\Big[1+\frac{d-3}{d-2}\kappa^2\lambda^2\Big]z^{d-1}+\frac{d-3}{d-2}\kappa^2\lambda^2z^{2(d-2)}~.
\end{equation}

\subsection*{Analysis by matching method}
Now we proceed to find the relationship between the critical temperature $T_c$ and charge density $\rho$ using the matching method \cite{Gregory et. al. 01}. For that we expand $\phi(z)$ and $\psi(z)$ in Taylor's series around $z=1$ and equate it with the boundary condition mentioned in eq.(s)(\ref{Bc 1}, \ref{Bc2})  at some point $z=z_m$. This yields
\begin{equation}
\label{phi mathing 1}
\Big[\mu-\frac{\rho}{r_+^{d-3}}z^{d-3}\Big]_{z=z_m}=\Big[\phi(1)-(1-z)\phi'(1)+\frac{1}{2}(1-z)^2\phi''(1)-\mathcal{O}((1-z)^3)\Big]_{z=z_m}~.
\end{equation}
Now using the values of $\chi'(z)$ and $\chi(z)$ at the critical temperature given in eq.(\ref{crtcl temp 1}) and eq.(\ref{crtcl temp 2}), in  eq.(\ref{1}), we end up with

\begin{equation}
\label{phi matching 2}
\phi''(z)-\frac{d-4}{z}\phi'(z)-\frac{2r_+^2\phi(z)\psi^2(z)}{f(z)z^4}=0~.
\end{equation}
From the above equation, we get 

\begin{equation}
\label{phi matching 3}
\phi''(1)=\Bigg[(d-4)+\frac{2\psi^2(1)}{g'_0(1)}\Bigg]\phi'(1)~.
\end{equation}
With this result eq.(\ref{phi mathing 1}) simplifies to the form

\begin{equation}
\label{phi matching 4}
\Bigg\{\mu-\frac{\rho}{r_+^{d-3}}z^{d-3}=-(1-z)\phi^{'}(1)+ \frac{1}{2}(1-z)^{2}\Bigg((d-4)+\frac{2\psi^2(1)}{g'_0(1)}\Bigg)\phi'(1)\Bigg\}_{z=z_m}~.
\end{equation} 

\noindent
Taking derivative on both sides of the above relation and setting $z=z_m$ gives

\begin{equation}
\label{phi matching 5}
\Bigg\{-(d-3)\frac{\rho}{r_+^{d-3}}z^{d-4}=\phi'(1)-(1-z)\Bigg((d-4)+\frac{2\psi^2(1)}{g'_0(1)}\Bigg)\phi'(1)\Bigg\}_{z=z_m}~.
\end{equation}
Setting  $\psi(1)=\alpha,~\phi'(1)=-v$ and $\tilde{v}=\frac{v}{r_+}$ in the above equation, we arrive at
\begin{equation}
\label{phi mtchng 6}
\alpha^2=\frac{g_0'(1)}{2(1-z_m)}\Big[1-(1-z_m)(d-4)\Big]\Bigg[1-\Big(\frac{T_c}{T}\Big)^{d-2}\Bigg]
\end{equation}
where

\begin{equation}
\label{phi mtchng 7}
T_c=\xi\rho^{\frac{1}{d-2}}    
\end{equation}

\begin{equation}
\label{xi}
\xi=-\Big(\frac{g_0'(1)}{4\pi}\Big)\frac{(d-3)^{\frac{1}{d-2}}z_m^{\frac{d-4}{d-2}}}{\tilde{v}^{\frac{1}{d-2}}\big[1-(1-z_m)(d-4)\big]^{\frac{1}{d-2}}}~.
\end{equation}

\noindent Now we want to figure out the form of $\tilde{v}$ in terms of known parameters $z_m$, $d$, $m^2$. Here again we use eq.(s) (\ref{crtcl temp 1}, \ref{crtcl temp 2}) in eq.(\ref{Scalar EoM A}) and end up getting

\begin{equation}
\label{psi matching 1}
\psi''(z)+\Big[\frac{f'(z)}{f(z)}-\frac{d-4}{z}\Big]\psi'(z)+\frac{r_+^2}{z^4}\Big[\frac{\phi^2(z)}{f^2(z)}-\frac{m^2}{f(z)}\Big]\psi(z)=0~.
\end{equation}
From this relation, we get

\begin{gather}
\label{psi matching 2}
\psi'(1)=\frac{m^2}{g_0'(1)\psi(1)} \\
\label{psi matching 3}
\psi''(1)=\frac{1}{2}\Big[(d-4)-\frac{g_0''(1)}{g_0'(1)}+\frac{m^2}{g_0'(1)}\Big]\frac{m^2}{g_0'(1)}\psi(1)-\frac{{\phi'}^2(1)}{2r_+^2{g_0'}^2(1)}\psi(1)~.
\end{gather}
The Taylor's series expansion of $\psi(z)$ around $z=1$ reads

\begin{equation}
\label{psi matching 4}
\psi(z)=\psi(1)-(1-z)\psi'(1)+\frac{(1-z)^2}{2!}\psi''(1)+....~.
\end{equation}
Following the matching technique as earlier, gives the expression for $\tilde{v}$ to be   

\begin{equation}
\label{vtilda}
\tilde{v}^2=m^4+m^2g_0'(1)\Big[d-4-\frac{g_0''}{g_0'}\Big]-\frac{4m^2g_0'\big[\Delta_+(1-z_m)+z_m\big]}{(1-z_m)\big[\Delta_+(1-z_m+2z_m)\big]}+\frac{4{g_0'}^2\Delta_+}{(1-z_m)\big[\Delta_+(1-z_m)+2z_m\big]}~.
\end{equation}

\noindent The values of  $g_{0}^{'}(1)$, $g_{0}^{''}(1)$ and $\xi$ have been calculated from eq.(\ref{gz}) and eq.(\ref{xi}) respectively. For that we have taken different values of $\lambda$ obtained using Sturm-Liouville eigen value method corresponding to different back-reaction parameter $\kappa$ \cite{Debabrata-EPJC} . The values of $\lambda$ and $\xi$ are displayed in Table [\ref{k-lambda}] for different back-reaction parameter $\kappa$. In the calculations, we have chosen $m^2=-3$ and $d=5$.

\begin{table}[ht]
	\caption{Values of $\lambda$ and $\xi$ for different $\kappa$ [ $m^{2}=-3$, $d=5$ ] }   
	\centering                          
	\begin{tabular}{|c| c| c|}            
		\hline
		$\kappa$ & $\lambda$ & $\xi$  \\
		\hline
		0 & 18.23 & 0.2016 \\
		\hline
		0.05 & 18.11 & 0.1997  \\
		\hline
		0.10 & 17.75 & 0.1938  \\
		\hline
		0.15& 17.16 & 0.1839   \\
		\hline
	\end{tabular}
	\label{k-lambda}  
\end{table}

\begin{figure}[h!]
	\centering
	\includegraphics[width=10cm]{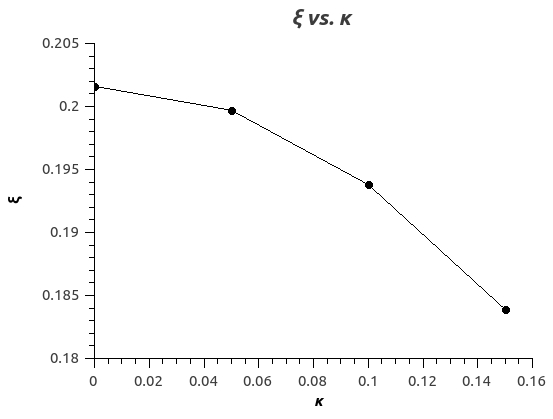}
	\caption{$\xi$ vs $\kappa$ plot : $z_{m}=0.5, m^2=-3, d=5$}
	\label{xi_kappa}
\end{figure}

In Fig[\ref{xi_kappa}], we have plotted the results of  $\xi$ vs. $\kappa$. The plot shows that the critical temperature decreases with increase in the back-reaction parameter.


\section{Condensate with magnetic field}

In this section we shall explore the behaviour of the condensate solution in the presence of a magnetic field. To make progress we proceed with the ansatz
\begin{equation}
\label{Gauge-Scalar-ansatz-1}
A_\mu=(\phi(r),~0,~Bx,~0,~0,~...),~\psi\equiv\psi(r,~x)~.
\end{equation}
With this choice in hand, we rewrite  eq.(\ref{Scalar EoM}) as
\begin{eqnarray}
\label{Scalar EoM B}
r^2f(r)\Bigg[\frac{\partial^2\psi(r,x)}{\partial r^2}+\Big(\frac{d-2}{r}+\frac{f'(r)}{f(r)}-\frac{\chi'(r)}{2}\Big)\frac{\partial \psi(r,x)}{\partial r}+\Big(\frac{q^2e^{\chi(r)}}{f^2(r)}\phi^2-\frac{m^2}{f(r)}\Big)\psi(r,x)\Bigg] \nonumber\\
=-\frac{\partial^2 \psi(r,x)}{\partial x^2}+q^2B^2x^2\psi(r,x)~.
\end{eqnarray}
Using the separation of variables method, we have
\begin{equation}
\label{SoV}
\psi(r,x)=R(r)X(x)~.
\end{equation}
Eq.(\ref{Scalar EoM B}) now decouples into two differential equations which has the following forms
\begin{subequations}
	\label{Decoupled EoM}
	\begin{gather}
	\label{1st}
	\frac{1}{X(x)}\frac{d^2 X(x)}{d x^2}-q^2B^2x^2=-K \\
	\label{2nd}
	\frac{r^2f(r)}{R(r)}\Bigg[\frac{\partial^2R(r)}{\partial r^2}+\Big(\frac{d-2}{r}+\frac{f'(r)}{f(r)}-\frac{\chi'(r)}{2}\Big)\frac{\partial R(r)}{\partial r}+\Big(\frac{q^2e^{\chi(r)}}{f^2(r)}\phi^2-\frac{m^2}{f(r)}\Big)R(r)\Bigg]=K
	\end{gather}
\end{subequations}
where $K$ is the constant of separation. Eq.(\ref{1st}) is of the form of Schrodinger's equation for a simple harmonic oscillator. The eigen values are $K=\lambda_n q B$ with $\lambda_n=2n+1$. The corresponding eigenfunctions can be given in terms of Hermite polynomials, $H_n(\sqrt{2qB}x)$. As explained in \cite{Albash et. al. 01}, for the rest of our analysis we shall work with the lowest mode ($n=0$) solution,  which implies $\lambda_0=1$. Eq.(\ref{2nd}) can now be written as  (introducing new coordinate variable $z=\frac{r_+}{r}$)
\begin{equation}
\label{2ndz}
R''(z)+\Bigg[\frac{f'(z)}{f(z)}-\frac{d-4}{z}-\frac{\chi'(z)}{2}\Bigg]R'(z)+\Bigg[\frac{e^{\chi(z)}r_+^2\phi^2(z)}{z^4f^2(z)}-\frac{m^2r_+^2}{z^4f(z)}-\frac{B}{z^2f(z)}\Bigg]R(z)=0~.
\end{equation}
We shall employ the matching method once again, discussed in the previous section, to find out the relation between the critical magnetic field ($B_c$) and critical temperature ($T_c$). For that, first we make a Taylor's series expansion of $R(z)$ around $z=1$ which reads
\begin{equation}
R(z)=R(1)-R^{'}(1)(1-z)+\frac{1}{2}R^{''}(1)(1-z)^{2}+O\left((1-z)^{3}\right)~.
\label{Taylor}
\end{equation}
Further the asymptotic form for $R(z)$ reads
\begin{equation}
R(z)=\frac{\left \langle O \right \rangle_{+}}{r_{+}^{\lambda_{+}}}z^{\lambda_{+}}~.
\label{Asymp}
\end{equation}
At any interior point $z=z_m$ we match these solutions. Hence, equating these at $z=z_{m}$ yields
\begin{equation}
\left[\frac{\left \langle O \right \rangle_{+}}{r_{+}^{\lambda_{+}}}z^{\lambda_{+}}\right]_{z=z_{m}}=\left[R(1)-R^{'}(1)(1-z)+\frac{1}{2}R^{''}(1)(1-z)^{2}+O\left((1-z)^{3}\right)\right]_{z=z_{m}}
\label{eq55}~.
\end{equation}
Differentiating eq.(s)(\ref{Taylor}, \ref{Asymp}) with respect to $z$ and evaluating at $z=z_{m}$ yields
\begin{equation}
\left[\lambda_{+}\frac{\left \langle O \right \rangle_{+}}{r_{+}^{\lambda_{+}}}z^{\lambda_{+}-1}\right]_{z=z_{m}}
=\left[R^{'}(1)-R{''}(1)(1-z)+O\left((1-z)^{3}\right)\right]_{z=z_{m}}~.
\label{eq56}
\end{equation}
As was discussed in the earlier section, near the critical temperature we put $\chi'(z)=\chi(z)=0$ in eq.(\ref{2ndz}). This simplifies the form of eq.(\ref{2ndz}) and it reads
\begin{equation}
R^{''}(z)+\left(\frac{f^{'}(z)}{f(z)}-\frac{d-4}{z}\right)R^{'}(z)+
\frac{\phi^{2}(z)r_{+}^{2}R(z)}{z^{4}f^{2}(z)}-\frac{m^{2}r_{+}^{2}R(z)}{z^{4}f(z)}\\=\frac{BR(z)}{z^{2}f(z)}~.
\label{eq52}
\end{equation}
From this equation, we obtain
\begin{equation}
R^{'}(1)=\left(\frac{m^{2}}{g_{0}^{'}(1)}+\frac{B}{r_{+}^{2}g_{0}^{'}(1)}\right)R(1)
\label{eq57}
\end{equation}

\begin{multline}
R^{''}(1)=
\frac{1}{2}\left[d-4+\frac{m^{2}}{g_{0}^{'}(1)}+\frac{B}{r_{+}^{2}g_{0}^{'}(1)}-\frac{g_{0}^{''}(1)}{g_{0}^{'}(1)}\right]\left[\frac{m^{2}}{g_{0}^{'}(1)}+\frac{B}{r_{+}^{2}g_{0}^{'}(1)}\right]R(1)
\\+\frac{BR(1)}{r_{+}^{2}g_{0}^{'}(1)}
-\frac{\phi^{'2}(1)R(1)}{2r_{+}^{2}g_{0}^{'2}(1)}~.
\label{eq58}
\end{multline}

\noindent Substituting $R^{'}(1)$ and $R^{''}(1)$ in eq.(s)(\ref{eq55}, \ref{eq56}), we have

\begin{multline}
\left[\frac{\left \langle O \right \rangle_{+}}{r_{+}^{\lambda_{+}}}z_{m}^{\lambda_{+}}\right]=
R(1)
-\left(\frac{m^{2}}{g_{0}^{'}(1)}+\frac{B}{r_{+}^{2}g_{0}^{'}(1)}\right)(1-z_{m})R(1)\\
+\frac{1}{2}(1-z_{m})^{2}
[\frac{1}{2}\left[d-4+\frac{m^{2}}{g_{0}^{'}(1)}+\frac{B}{r_{+}^{2}g_{0}^{'}(1)}-\frac{g_{0}^{''}(1)}{g_{0}^{'}(1)}\right]\left[\frac{m^{2}}{g_{0}^{'}(1)}+\frac{B}{r_{+}^{2}g_{0}^{'}(1)}\right]R(1)
\\+\frac{BR(1)}{r_{+}^{2}g_{0}^{'}(1)}
-\frac{\phi^{'2}(1)R(1)}{2r_{+}^{2}g_{0}^{'2}(1)}]
\label{eq59}
\end{multline}

\begin{multline}
\left[\lambda_{+}\frac{\left \langle O \right \rangle_{+}}{r_{+}^{\lambda_{+}}}z_{m}^{\lambda_{+}-1}\right]=
\left(\frac{m^{2}}{g_{0}^{'}(1)}+\frac{B}{r_{+}^{2}g_{0}^{'}(1)}\right)R(1)\\
-(1-z_{m})
[\frac{1}{2}\left[d-4+\frac{m^{2}}{g_{0}^{'}(1)}+\frac{B}{r_{+}^{2}g_{0}^{'}(1)}-\frac{g_{0}^{''}(1)}{g_{0}^{'}(1)}\right]\left[\frac{m^{2}}{g_{0}^{'}(1)}+\frac{B}{r_{+}^{2}g_{0}^{'}(1)}\right]R(1)
\\+\frac{BR(1)}{r_{+}^{2}g_{0}^{'}(1)}
-\frac{\phi^{'2}(1)R(1)}{2r_{+}^{2}g_{0}^{'2}(1)}]~.
\label{eq60}
\end{multline}

\noindent Eq.(s)(\ref{eq59}) and (\ref{eq60}) yields a quadratic equation for $B$. This reads
\begin{equation}
B^{2}+pr_{+}^{2}B+nr_{+}^{4}-\phi^{'2}(1)r_{+}^{2}=0
\label{critical mag}
\end{equation}
where

\begin{equation}
p=2m^{2}+\left(d-4-\frac{g^{''}_{0}(1)}{g^{'}_{0}(1)}\right)g_{0}^{'}(1)+2g_{0}^{'}(1)-\frac{4g_{0}^{'}(1)(\lambda_{+}(1-z_{m})+z_{m})}{(1-z_{m})(\lambda_{+}(1-z_{m})+2z_{m})}
\label{eq62}
\end{equation}
and


\begin{multline}
n=m^{4}
+m^{2}g_{0}^{'}(1)
\left[\left(d-4-\frac{g^{''}_{0}(1)}{g^{'}_{0}(1)}\right)
-\frac{4(z_{m}+\lambda_{+}(1-z_{m}))}{(1-z_{m})(2z_{m}+\lambda_{+}(1-z_{m}))}\right]\\
+\frac{4\lambda{+}g_{0}^{'2}(1)}{(1-z_{m})(2z_{m}+\lambda_{+}(1-z_{m}))}~.
\label{eq63}
\end{multline}

\noindent Since at $T=T_{c}$, the scalar field vanishes ($\psi(z)=0$), this gives $\phi^{'}(1)$ from eq.(\ref{sol. 1}) and reads
\begin{equation}
\phi^{'}(1)= -\lambda r_{+} (d-3)~.
\end{equation}
\noindent Substituting this in eq.(\ref{critical mag}) gives the critical magnetic field to be

\begin{eqnarray}
B_{c}=\bigg(\frac{1}{2}\bigg)\bigg(-\frac{g_{0}^{'}(1)}{4\pi}\bigg)^{d-4}\bigg(\frac{1}{\xi}\bigg)^{d-2}\bigg[\Omega(d,m,z_{m})-p\bigg(-\frac{4\pi\xi}{g_{0}^{'}(1)}\bigg)^{d-2}\bigg(\frac{T}{T_{c}}\bigg)^{d-2}\bigg]
\end{eqnarray}
where
\begin{equation}
\Omega(d,m,z_{m})=\bigg[4(d-3)^{2}-(4n-p^{2})\bigg(-\frac{4\pi\xi}{g_{0}^{'}(1)}\bigg)^{(2d-4)}\bigg(\frac{T}{T_{c}}\bigg)^{(2d-4)}\bigg]~.
\end{equation}

\begin{figure}[h!]
	\centering
	\includegraphics[width=10cm]{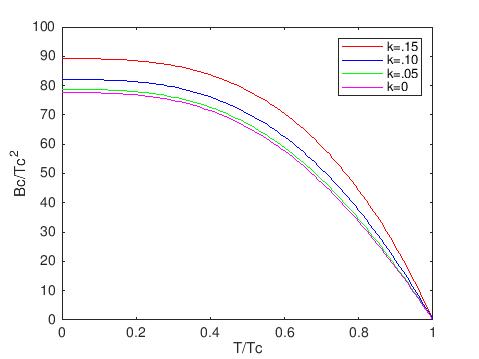}
	
	\caption{$B_{c}/T_{c}^{2}$ vs $T/T_{c}$ plot : $z_{m}=0.5, m^2=-3, d=5$}
	\label{Bc_T}
\end{figure}

\noindent In Fig.[\ref{Bc_T}], we have plotted $B_{c}/T_{c}^2$ against $T/T_{c}$ for different back-reaction parameters $\kappa$, with $z_m= 0.5$, $m^2=-3$, $d= 5$. It is evident from the figure that there exists a certain critical temperature $T_c$ and a critical magnetic field $B_c$ above which the superconducting phase vanishes.

\noindent From Fig.[2], it is to be noted that the ratio of $\frac{B_c}{T_c^2}$ at $T=0$ increases with increase in the back reaction parameter. The value of $B_c$ can be estimated from the ratio $\frac{B_c}{T_c^2}$ and using eq.(\ref{phi mtchng 7}). These values have been displayed in Table [2]. 

\begin{table}[ht]
	\caption{Values of $B_{c}$ for different $\kappa$ [ $m^{2}=-3$, $d=5$, $T=0$ ] }   
	\centering                          
	\begin{tabular}{|c| c| c|}            
		\hline
		$\kappa$ & $\frac{B_{c}}{T_{c}^{2}}$ & $B_{c}$  \\
		\hline
		0 & 77.5879 & 3.1533 $\rho^{\frac{2}{3}}$ \\
		\hline
		0.05 & 78.6834 & 3.1379 $\rho^{\frac{2}{3}}$ \\
		\hline
		0.10 & 82.0542 & 3.0818$\rho^{\frac{2}{3}}$ \\
		\hline
		0.15& 89.1723 & 3.0157$\rho^{\frac{2}{3}}$  \\
		\hline
	\end{tabular}
	\label{k-B_{c}}  
\end{table}

\noindent From Table [2], it can be observed that the critical magnetic field $B_c$ decreases with increase in the back reaction parameter $\kappa$. This indicates that the presence of the back reaction parameter destroys the superconducting phase earlier, that is, for a smaller value of the critical magnetic field $B_c$.  

\section{Conclusion}
In this paper we have studied the influence of back reaction on holographic superconductor in the framework of Maxwell electrodynamics using the matching method. The holographic superconductor model that we have considered in our analysis consists of Einstein-Hilbert gravity theory along with a complex scalar field minimally coupled to Maxwell field. Further we have investigated the effect of an external magnetic field in our holographic superconductor model. From our analysis we can conclude from the critical temperature and charge density relationship that the critical temperature depends on both charge density and the back reaction parameter. We have presented the analytical results for the ratio of the critical temperature and charge density for $d=5$ and $m^2=-3$. Our results indicate that condensation gets harder to form as we include the effect of back reaction. With these findings, we then obtain the expression for the critical magnetic field above which the superconducting phase vanishes. This is obtained using matching method once again. We observe that the ratio of $B_c$ and $T_{c}^{2}$ at $T=0$ increases with the increase in back reaction parameter $\kappa$. However, we have found that the critical magnetic field $B_c$ decreases with increase in the back reaction parameter $\kappa$. This clearly tells that the presence of the back reaction parameter destroys the superconducting phase for a smaller value of the critical magnetic field $B_c$.

\section*{Acknowledgment}
S. Pal wants to thank the Council of Scientific and Industrial Research (CSIR), Govt. of India for financial
support. S. Gangopadhyay  acknowledges the support of IUCAA, Pune for the Visiting Associateship.


\end{document}